\documentclass[letterpaper, 10 pt, conference]{ieeeconf}  

\IEEEoverridecommandlockouts                              

\usepackage[sorting=none, style=ieee,doi=false,isbn=false,url=false,eprint=false]{biblatex}
\usepackage{amsmath,amssymb,amsfonts}
\usepackage{algorithmic}
\usepackage{romannum}
\usepackage{color}
\usepackage{comment}
\usepackage{dirtytalk}
\usepackage{theorem}
\usepackage[capitalise]{cleveref}
\usepackage{siunitx}
\usepackage{bbm}

\usepackage{tikz,pgfplots}
\usepackage{float}
\usepackage{algorithm}
\usepackage{algorithmic}
\usepackage{etoolbox}
\usepackage{graphicx}
\usepackage{caption}
\usepackage{subcaption}
\usepackage{multicol}
\usepackage{lipsum}
\usepackage{pgf-pie}
\usepackage[raggedright]{sidecap}
\usepackage{capt-of}
\usepackage{xspace}
\usepackage{titlesec}
\usepackage{multicol}

\newtheorem{theorem}{Theorem}[section]
\newtheorem{definition}{Definition}[section]

\newtheorem{corollary}{Corollary}[section]
\newtheorem{proposition}{Proposition}[section]

\titlespacing\section{0pt}{2pt plus 1pt minus 1pt}{1pt plus 1pt minus 1pt}
\titlespacing\subsection{0pt}{2pt plus 1pt minus 1pt}{2pt plus 1pt minus 1pt}

\title{\LARGE \bf
A Safe Control Architecture Based on Robust Model Predictive Control for Autonomous Driving*
}

\author{Maryam Nezami$^{1}$, Ngoc Thinh Nguyen$^{2}$, Georg Männel$^{3,1}$, Hossam Seddik Abbas$^{1}$ and Georg Schildbach$^{1}$ 
\thanks{$^{1}$Institute for Electrical Engineering in Medicine,
        University of Lübeck, Lübeck, Germany
        {\tt\small \{maryam.nezami, ge.maennel, h.abbas, georg.schildbach\}@uni-luebeck.de}}
\thanks{$^{2}$Institute for Robotics and Cognitive Systems, University of Lübeck, Lübeck, Germany
        {\tt\small nguyen@rob.uni-luebeck.de}}
\thanks{$^{3}$Fraunhofer Research Institution for Individualized and Cell-Based Medical Engineering, Fraunhofer IMTE, Lübeck, Germany
        {\tt\small georg.maennel@imte.fraunhofer.de}}
 \thanks{*N.T. Nguyen is funded by the German Ministry of Food and Agriculture (BMEL) Project No. 28DK133A20. 
 G. Männel is funded by the European Union - European Regional Development Fund (ERDF), the Federal Government and Land Schleswig Holstein Project No. 12420002. H. S. Abbas is funded by the German Research Foundation (DFG), project number 419290163.}%
}

\begin{document}

\maketitle
\thispagestyle{empty}
\pagestyle{empty}

\begin{abstract}
This paper proposes a Robust Safe Control Architecture (RSCA) for safe-decision making. The system to be controlled is a vehicle in the presence of bounded disturbances. The RSCA consists of two parts: a Supervisor MPC and a Controller MPC. Both the Supervisor and the Controller are tube MPCs (TMPCs). The Supervisor MPC provides a safety certificate for an operating controller and a backup control input in every step. After an unsafe action by the operating controller is predicted, the Controller MPC takes over the system. In this paper, a method for the computation of a terminal set is proposed, which is robust against changes in the road curvature and forces the vehicle to reach a safe reference. Moreover, two important proofs are provided in this paper. First, it is shown that the backup control input is safe to be applied to the system to lead the vehicle to a safe state. Next, the recursive feasibility of the RSCA is proven. By simulating some obstacle avoidance scenarios, the effectiveness of the proposed RSCA is confirmed.
\end{abstract}


\section{INTRODUCTION}

Advances in technology have brought more attention to autonomous vehicles. However, autonomous vehicles are among safety-critical systems, i.e., their failure can cause harm to human health, life, property, or the environment~\cite{Knight2002}. For this reason, despite rapid advancements in Artificial Intelligence (AI) and Machine Learning (ML), control methods based on AI and ML usually cannot directly be applied to autonomous vehicles. A possible solution is a Safe Control Architecture (SCA) that provides safety certificates for any control input, regardless of the \textit{operating controller} which generates the control input, e.g.~\cite{Nezami2021, Akametalu2014,Wabersich2018}. The \textit{operating controller} refers to a potentially unsafe control algorithm that can be treated as a black box. It could be an AI or ML based, a trajectory tracking algorithm, or even a human driver. 

There is a vast literature on approaches for the safe decision-making of autonomous vehicles, e.g.~\cite{minderhoud2001,abbink2019, gao2020,koschi2020,Tearle2021,Gray2013}. In~\cite{minderhoud2001}, time-to-collision (TTC) is calculated and intervention happens if the car passes the TTC threshold. In~\cite{abbink2019}, an adaptive controller is proposed based on time to lane crossing to shift the authority between the driver and the controller. In~\cite{gao2020}, under uncertain conditions, the probability of collision between multiple vehicles is estimated to provide a situational assessment. The method in~\cite{koschi2020} is based on a set-based prediction of the behavior of all traffic participants. In~\cite{Tearle2021}, a safety filter is proposed based on solving an MPC problem to maintain the car within the road boundaries. In~\cite{Gray2013}, safe decision-making is attained by enforcing safety constraints in an MPC problem. 

A SCA for safe decision-making, using a \textit{supervisory controller} that continuously provides a safety certificate for the control actions of the operating controller, is proposed in~\cite{Nezami2021}. In~\cite{Nezami2021} the supervisory controller consists of a Supervisor Model Predictive Control (MPC). The safety certificate is provided by predicting the system state one step ahead and based on the feasibility status of an optimization problem.
However, the SCA only uses a linearized nominal model, not considering model uncertainties or disturbances that might act on the system. Hence, due to wind or a change in surface friction, wrong decisions by the SCA are possible. 
Therefore in this paper, we propose a robust SCA (RSCA) based on tube MPC (TMPC) that can provide a safety certificate for an arbitrary controller for safe decision-making even in the presence of disturbances. Furthermore, the proposed RSCA provides another TMPC which replaces the \emph{operating controller} once it is considered unsafe.


\emph{Contributions:} This paper makes four main contributions. First, an RSCA based on TMPC is proposed. The RSCA is based on tightening the state and input constraints in optimization problem such that the architecture can guarantee the safety of the system even when it is disturbed.
Second, a method for computing a terminal set that is robust against the changes in the road curvature is proposed. Third, the recursive feasibility of the proposed RSCA is established. Finally, the validation of the proposed RSCA is shown in a case study on a safe obstacle avoidance scenario by a vehicle. 


\emph{Content:} Section II reviews the concept of the Supervisor MPC from~\cite{Nezami2021}. Section III describes the model that represents the vehicle in this paper. Section IV explains the RSCA and the setup of the TMPCs. Section V presents the method for the computation of the terminal set and the theoretical safety guarantees. Section VI demonstrates the simulation setup and the results. Section VII presents the conclusion. 

\emph{Notation:} In this paper the symbols $\oplus$ and $\ominus$ denote Minkowski sum and Pontryagin difference, respectively. The function $\text{diag}(\mathbf{x})$ constructs a diagonal matrix from a vector $\mathbf{x}$ and $\boldsymbol{1}$ denotes a vector of ones of suitable dimension.









\section{concept of Supervisor MPC}
Consider the discrete-time nonlinear system
\begin{equation}
	x_{k+1} = f(x_k,u_k),\quad x_0 = \bar{x}_0, \quad \forall k \in \mathbb{Z}_{0+},
	\label{nonlinear_model}  
\end{equation} 
where, $\bar{x}_0 \in \mathbb{R}^n$, $x_k \in \mathbb{R}^n$ and $u_k \in \mathbb{R}^m$ are the initial condition, the state vector and the input vector at time step $k$, respectively. System~\eqref{nonlinear_model} is subject to polytopic state constraints, $x_k \in \mathcal{X}_k$, and polytopic input constraints, $u_k \in \mathcal{U}$. The control input generated by the operating controller, whose safety should be certified, is $u^{\rm o}_{k} \in \mathbb{R}^m$. 
\vspace{-4mm}
\begin{definition}[\cite{Nezami2021}]
		\normalfont{An \textbf{\textit{operating controller}} is an arbitrary control algorithm that generates the control inputs $u^{\rm o}_{k} \in \mathbb{R}^m$ to fulfill a control goal for system~\eqref{nonlinear_model}}.
	\end{definition}
\vspace*{-5mm}
\begin{definition}[\cite{Nezami2021}]
		\normalfont{ At any time step $k$, system (\ref{nonlinear_model}) is in a \textbf{\textit{safe state}} if there exists a feasible control sequence such that the input and state constraints are satisfied at all time steps in the future, i.e., $\exists \{ u_k , u_{k+1}, \cdots \} \quad \text{such that} \quad x_{k+j} \in \mathcal{X}_{k+j}, u_{k+j} \in \mathcal{U}, \forall j= 0,1,2, \cdots$.}
\end{definition}
\vspace*{-5mm}
\begin{definition}[\cite{Nezami2021}]
		\normalfont{A \textbf{\textit{safety event}} is when the application of $u^{\rm o}_{k}$ drives the system~\eqref{nonlinear_model} from a safe state to an unsafe state}.
\end{definition}
Just the detection of a safety event is not enough, because some time for a reaction is required to intervene with $u^{\rm o}_{k}$. Therefore, the concept of a \textbf{\textit{detection event}} is introduced, which is the prediction of a \emph{safety event}. In the SCA, using an MPC as the Supervisor to predict a safety event is suggested. \say{An \emph{MPC-based detection event} is the fact that the following optimization problem is infeasible~\cite{Nezami2021}}:
\begin{subequations}\label{MPC_previous_paper}
	\begin{align} 
		\underset{U}{\text{min}}
		&\sum_{i=1}^{N} x^\top_{i|k} Q x_{i|k} + u^\top_{i-1|k} R u_{i-1|k}  \\
		 \text{s.t.} \;\;
		& x_{i|k} = f(x_{i-1|k},u_{i-1|k}), \quad \forall i = 1,\cdots,N,  \label{model_in_PMPC} \\
		& x_{0|k} = \hat{x}_{k+1}, \label{intial_condition_in_PMPC} \\
		&  x_{i|k} \in \mathcal{X}_{i|k}, \quad \forall i = 1,2,\cdots,N,  \label{state_constraint_in_PMPC}\\
		&  u_{i-1|k} \in \mathcal{U}, \quad \forall i = 1,2,\cdots,N, \label{input_contsraint_in_PMPC}
	\end{align}
\end{subequations}
where, $Q \succeq 0 \in \mathbb{R}^{n\times n}$ and $R \succ 0 \in \mathbb{R}^{m\times m}$ are tuning matrices. The initial condition is $x_{0|k} \in \mathbb{R}^n$, whose value in step $k$ of the optimization problem is $\hat{x}_{k+1} = f(x_k,u^{\rm o}_{k})$, where $x_k \in \mathbb{R}^{n}$ is the system's current state. The prediction horizon is $N$ and $U = \{ \bar{u}_{0|k}, \cdots, \bar{u}_{N-1|k} \}$ is the decision variable. The state and input constraints are enforced in the optimization problem in~\eqref{state_constraint_in_PMPC} and~\eqref{input_contsraint_in_PMPC}, respectively. 

 The infeasibility of the optimization problem~\eqref{MPC_previous_paper} means that if $u^{\rm o}_{k}$ is applied to the system~\eqref{nonlinear_model} and the system goes to $\hat{x}_{k+1}$, then there will be no feasible solution for the optimization problem~\eqref{MPC_previous_paper} with respect to the state and input constraints in the next step. In this case, $u^{\rm o}_{k}$ is labeled as unsafe and should not be applied to the system (detection event). What happens after a detection event depends on the application. In our case, we assume that after the \textit{operating controller} makes a mistake, the system is controlled by another MPC.



\section{vehicle model}
In the remainder of this paper, it is assumed that the system is a car that is moving at a constant longitudinal speed and that we can measure the system states in every step. Then, the vehicle's lateral dynamics can be represented in terms of error with respect to the road as follows~\cite[p.~36]{Rajamani2012}: 
\begin{multline}
	\!\!\!\!\!\!\!\!\frac{\text{d}}{\text{d}t}\!\!\begin{bmatrix}
		\!e_{\rm y}(t)\!\! \\ \!\dot{e}_{\rm y}(t)\!\!\\ \!e_\psi(t) \! \!\\ \!\dot{e}_\psi(t)\! \!
	\end{bmatrix}\!\!\! = \!
\!\!\begin{bmatrix}
			0    \!\!  &  \!\!   1  \!\!   &    \!\! 0   \!\!  &\!\!    0     \\
			0  \!\!    & \!\!    a \!\!    &   \!\! b    \!\!     &  \!\!  c      \\
			0   \!\!   &   \!\!  0  \!\!   &  \!\!   0 \!\!    &  \!\!  1      \\
			0    \!\!  & \!\!    d  \!\!   &  \!\!  e   \!\!   &  \!\!  f       \\
	\end{bmatrix}\! \!\!\!\begin{bmatrix}
	\!	e_{\rm y}(t) \! \!\\ \!\dot{e}_{\rm y}(t)\! \! \\\! e_\psi (t)\! \! \\ \!\dot{e}_\psi(t) \! \!
	\end{bmatrix}
\!\!	+ \! \!\begin{bmatrix}
		\!	0 \! \\ \! \frac{2C_{\alpha f}}{m} \! \\
		\!	0 \! \\
		\!	\frac{2 C_{\alpha f} l_f}{I_z} \!\\
	\end{bmatrix}\!\!  \delta(t) \! + \!\begin{bmatrix}
		\!	0 \! \!\\ \! g \!\! \\ \! 0 \! \!\\\!  h \!\!
	\end{bmatrix}\!\! \dot{\psi}^{\rm{des}}(t),
	\label{Con_model}
\end{multline}
where, 

\begin{minipage}{0.25\textwidth}
\begin{equation*}
	\begin{split}
		a &= -\frac{2C_{\alpha f} + 2 C_{\alpha r}}{m V_x}, \\
		c &= \frac{-2 C_{\alpha f} l_f + 2 C_{\alpha r} l_r}{m V_x},  \\
		e &=  \frac{2 C_{\alpha f} l_f - 2 C_{\alpha r} l_r}{I_z},  \\
		g &= -\frac{2C_{\alpha f}l_f-2C_{\alpha r} l_r}{m V_x} - V_x,
	\end{split}
\end{equation*}
\vspace{0.008cm}
\end{minipage}
\begin{minipage}{0.2\textwidth}
\begin{equation*}
	\begin{split}
		b &= +\frac{2C_{\alpha f} + 2 C_{\alpha r}}{m}, \\
		d &= -\frac{2 C_{\alpha f} l_f - 2C_{\alpha r} l_r}{I_z V_x}, \\
		f &=  -\frac{2C_{\alpha f}l^2_f + 2 C_{\alpha r} l^2_r}{I_z V_x},  \\
		h &= -\frac{2 C_{\alpha f} l^2_f + 2 C_{\alpha r} l^2_r}{I_z V_x}.
	\end{split}
\end{equation*}
\vspace{0.008cm}
\end{minipage}
Here, $e_{\rm y}(t)$ and $e_\psi(t)$ are the distance of the center of gravity (CoG) of the vehicle from the center-line of the road and the orientation error of the vehicle with respect to the road, respectively. The rate of change of the road curvature is shown with $\dot{\psi}_{\rm{des}}(t)$ and it is calculated offline. The control input which should be calculated online is the front wheel steering angle, $\delta(t)$. \cref{para-table} shows the list of vehicle parameters, with their values and units.
\begin{table}
	\caption{Vehicle parameters used in vehicle modeling~\cite{Gottmann2018}}
	\label{para-table}
	\begin{center}
		\begin{tabular}{ c l c}
			\hline
			\textbf{Symbol} & \textbf{Parameter}& \textbf{Value}  \\
			\hline
			$C_{\alpha f}$ & Cornering Stiffness Front &  $153$\,kN$/$rad \\
			
			$C_{\alpha r}$ & Cornering Stiffness Rear & $191$ \,kN$/$rad \\
			
			$l_f$ & Distance CoG to Front Axle & $ 1.3$ \,m \\
			
			$l_r$ & Distance CoG to Rear Axle & $1.7$ \,m \\
			
			$I_z$ & Vehicle Yaw Inertia & $5250$ \,kgm$^2$ \\
			
			$V_x$ & Longitudinal Velocity of the Vehicle & $10$\,m$/$s \\
			
			$m$ & Vehicle Mass & $2500$\,kg \\
			
			$w$ & Vehicle Width & $1.8$\, \rm m\\
			\hline
		\end{tabular}
	\end{center}
\end{table}

Assuming all the signals to be piecewise constant, model~\eqref{Con_model} is discretized by using the exact discretization method~\cite{DeCarlo1989}, as follows:
\begin{equation}\label{dis_model}
x_{k+1} = A x_{k} + B u_{k} + E \dot{\psi}^{\rm{des}}_{k}.
\end{equation}
Model~\eqref{dis_model} in the presence of disturbances due to model mismatches and external physical disturbances is considered:
\begin{equation}\label{dis_model_with_disturbance}
	x_{k+1} = A x_{k} + B u_{k} + E \dot{\psi}^{\rm{des}}_{k} + d_k,
\end{equation}
where, $d_k \in \mathcal{D} \subset \mathbb{R}^4$ is a bounded disturbance in polytopic form containing the origin. Furthermore, we assume that the upper and lower bounds on $\dot{\psi}^{\rm{des}}_{k}$ are known, i.e., $\dot{\psi}^{\rm{des}}_{k} \in \Psi$, where $\Psi$ is an interval. The discrete system states are $e^{\rm y}_k$, $\Delta e^{\rm y}_k$, $e^{\psi}_k$ and $\Delta e^{\psi}_k$.

To keep the car within the road, the road boundaries are enforced in the problem by imposing constraints on $e^{\rm y}_k$, i.e., $ \lvert e^{\rm y}_k \rvert \leq \frac{R}{2} - \frac{w}{2}$, where $R$ is the road width and $w$ is the car width. The presence of an obstacle is enforced in the optimization problem by changing the constraint on $e^{\rm y}_k$. Moreover, the control input is bounded, i.e., $\lvert  u_{k} \rvert \leq \bar{\delta}$. Therefore, both the state constraint $\mathcal{X}_{k}$ and the input constraint $\mathcal{U}$ can be formulated in a polytopic form. 



\section{Robust safe control architecture}
\vspace{-0.1mm}
In this paper, a TMPC is used as the robust Supervisor. However, the generated tubes for tightening the state and input constraints in the RSCA cannot be the same as the standard TMPCs, e.g., in~\cite{Mayne2005}. The reason is that in those approaches, the TMPC is used as the controller of the system, and in every step, the current state of the system is available. Therefore, the error propagation in two steps should be considered in the computation of tubes. 

At step $k$, by using the system model~\eqref{dis_model}, the nominal state, denoted by $\hat{x}_{k+1}$, is calculated as follows:
\begin{equation}\label{estimated_state}
      \hat{x}_{k+1} = A x_k + B u^{\rm o}_{k} + E \dot{\psi}^{\rm{des}}_{k}, 
\end{equation}
Comparing~\eqref{dis_model_with_disturbance} and~\eqref{estimated_state}, the true state of the system is:
\begin{equation}\label{real_state}
    x_{k+1} = \hat{x}_{k+1} + d_k. 
\end{equation}

\begin{figure}
    \centering
    \includegraphics[scale = 0.3]{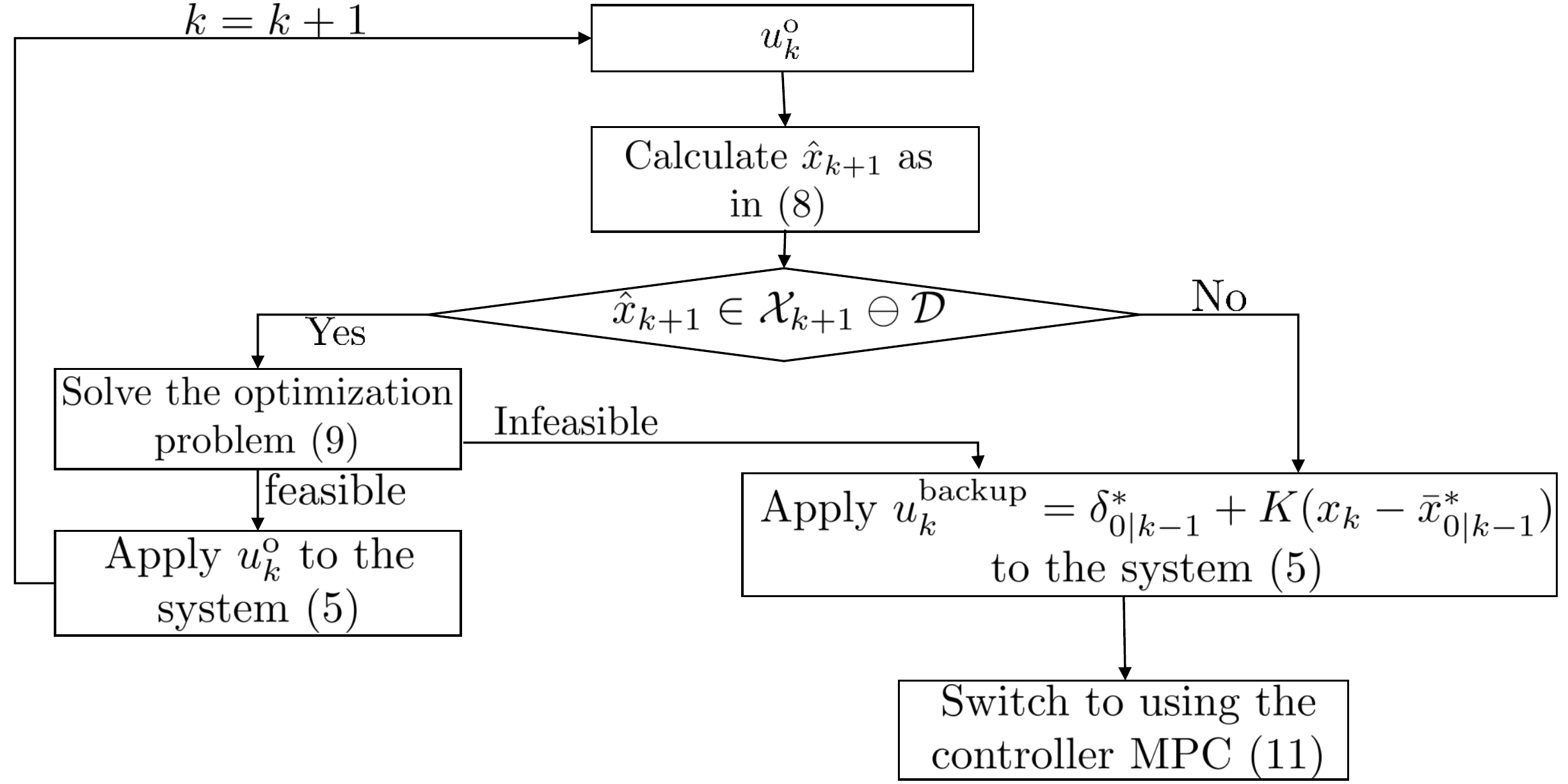}
    \caption{Flowchart of the proposed RSCA}
    \label{flow_chart}
\end{figure}

The flowchart of the proposed RSCA is shown in \cref{flow_chart}. 
If $\hat{x}_{k+1} \notin \mathcal{X}_{k+1} \ominus \mathcal{D}$, it is already concluded that $u^{\rm o}_{k}$ is unsafe, as $x_{k+1} \in \mathcal{X}_{k+1}$ cannot be guaranteed. If $\hat{x}_{k+1} \in \mathcal{X}_{k+1} \ominus \mathcal{D}$, still the safety of $u^{\rm o}_{k}$ is not approved because it has to be assured that the Supervisor MPC can provide a backup control input for the next step. The Supervisor MPC is as follows:
\begin{subequations}\label{RMPC}
	\begin{align} 
	\bar{J}\!\! =  &\min_{\bar{U},\bar{x}_{0|k}} \bar{x}^\top_{N|k} P \bar{x}_{N|k}\! + 
		\sum_{i=0}^{N-1}\!\! \bar{x}^\top_{i|k} Q \bar{x}_{i|k} \!\!+ R \bar{u}^2_{i|k}  \\
		 \text{s.t.} \;\;
		& \bar{x}_{i+1|k} \!= \!A \bar{x}_{i|k}\! + \! B \bar{u}_{i|k} \!+\! E \dot{\bar{\psi}}_{i|k}, \forall i \! =\! 0,\!\cdots\!,N\!\!-\!\!1,  \label{model_in_MPC} \\
		& \bar{x}_{i|k} \in \bar{\mathcal{X}}_{i|k} \ominus \mathcal{Z}, \quad \forall i \! = \! 0,\!\cdots\!,N-1,  \label{state_constraint_in_MPC}\\
		&  \bar{u}_{i|k} \in ( \mathcal{U} \ominus K \mathcal{Z} ) \ominus K\mathcal{D} , \quad \forall i \! = \!0,\!\cdots\!,N-1, \label{input_contsraint_in_MPC} \\
		& \bar{x}_{N|k} \in \mathcal{X}_N,  \label{terminal_cons_in_MPC} \\
    	& \hat{x}_{k+1} \in \bar{x}_{0|k} \oplus\mathcal{Z}, \label{intial_condition_in_MPC} 
	\end{align}
\end{subequations}
where, $\bar{x}_{0|k}$ and $\bar{U} = \{ \bar{u}_{0|k}, \cdots, \bar{u}_{N-1|k} \}$ are the decision variables. Model \eqref{model_in_MPC}, represents the nominal vehicle model where $\bar{x}_{i|k} \in \mathbb{R}^4$ and $\bar{u}_{i|k} \in \mathbb{R}$ are the predicted state and input at step $i+k+1$, and $\dot{\bar{\psi}}_{i|k}$ is defined as $\dot{\bar{\psi}}_{i|k} = \dot{\psi}^{\rm{des}}_{i+k+1}$. In the state constraint \eqref{state_constraint_in_MPC}, $\bar{\mathcal{X}}_{i|k}$ is defined as follows:
\begin{equation}
    \bar{\mathcal{X}}_{i|k} = \mathcal{X}_{i+k+1}  \label{x_constraint_in_SMPC}.
\end{equation}
The tuning matrices are  $Q \succeq 0 \in \mathbb{R}^{4\times 4}$ and $R > 0 \in \mathbb{R}$. The terminal set $\mathcal{X}_N \in \mathbb{R}^4$ is a robust positively invariant set for system \eqref{dis_model_with_disturbance}. The set $\mathcal{Z} \in \mathbb{R}^4$ is a disturbance invariant set for system \eqref{dis_model_with_disturbance} but without considering $E \dot{\psi}^{\rm{des}}_{k}$. The computation of $\mathcal{Z}$ and $\mathcal{X}_N$ is explained in the next section. The gain $K$ in~\eqref{input_contsraint_in_MPC} is chosen such that the matrix,
\begin{equation}\label{A_k}
    A_{\text{K}} = A+BK,
\end{equation}
is a stable matrix, where $A$ and $B$ are from~\eqref{dis_model_with_disturbance}.


As long as the optimization problem~\eqref{RMPC} is feasible, $u^{\rm o}_{k}$ is safe, i.e., $u_k = u^{\rm o}_{k}$. Also, the first element of the optimal control sequence $\bar{u}^*_{0|k}$ and $\bar{x}^*_{0|k}$ are to be saved in every step to be used in the calculation of a backup control input if necessary. At the moment, when a \emph{detection event} takes place, $u^{\rm o}_{k}$ is labeled as unsafe. Then, instead of $u^{\rm o}_{k}$ the following control input should be applied to the system~\eqref{dis_model_with_disturbance}:
\begin{equation}\label{take_over_control_input}
    u^{\text{backup}}_{k} = \bar{u}^*_{0|k-1} + K(x_k - \bar{x}^*_{0|k-1}),
\end{equation}
where, $x_k$ is the current state and $K$ is from~\eqref{A_k}.  


After a safety event is predicted, the following TMPC is suggested as the Controller MPC at the next steps:
\begin{subequations}\label{CMPC}
	\begin{align} 
\tilde{J} \!\!&=\!\! \min_{\tilde{U}, \Tilde{x}_{0|k}} \Tilde{x}^\top_{N\!-\!1|k} P \tilde{x}_{N\!-\!1|k}\! + \!
		\sum_{i=0}^{N-2}\!\! \Tilde{x}^\top_{i|k} Q \Tilde{x}_{i|k} \!\!+\! R \Tilde{u}^2_{i|k}  \\
		 \text{s.t.} \;\;
		& \Tilde{x}_{i+1|k} \!= \!A \Tilde{x}_{i|k}\! + \! B \Tilde{u}_{i|k} \!+\! E  \dot{\Tilde{\psi}}_{i|k}, \forall i \! =\! 0,\!\cdots\!,N\!\!-\!\!2,  \label{model_in_CMPC} \\
		& \Tilde{x}_{i|k} \in  \Tilde{\mathcal{X}}_{i|k} \ominus\mathcal{Z}, \quad \forall i \! = \! 0,\!\cdots\!,N-2,  \label{state_constraint_in_CMPC}\\
		&  \Tilde{u}_{i|k} \in \mathcal{U} \ominus K\mathcal{Z}, \quad \forall i \! = \!0,\!\cdots\!,N-2, \label{input_contsraint_in_CMPC} \\
		& \Tilde{x}_{N-1|k} \in \mathcal{X}_N,  \label{terminal_cons_in_CMPC} \\
    	& x_k \in \Tilde{x}_{0|k} \oplus\mathcal{Z}, \label{intial_condition_in_CMPC}
	\end{align}
\end{subequations}
where, $\tilde{U} = \{ \Tilde{u}_{0|k} , \cdots, \Tilde{u}_{N-2|k} \}$ and $\Tilde{x}_{0|k}$ are the decision variables. Model \eqref{model_in_CMPC}, represents the nominal vehicle model where $\Tilde{x}_{i|k} \in \mathbb{R}^4$ and $\Tilde{u}_{i|k} \in \mathbb{R}$ are the predicted state and input at step $k+i$, and $\dot{\Tilde{\psi}}_{i|k} = \dot{\psi}^{\rm{des}}_{i+k}$. In the state constraint~\eqref{state_constraint_in_CMPC}, $\Tilde{\mathcal{X}}_{i|k}$ is defined as follows:
\begin{equation}
    \Tilde{\mathcal{X}}_{i|k} = \mathcal{X}_{i+k}.  \label{x_constraint_in_CMPC}
\end{equation} 
The horizon of the Controller MPC~\eqref{CMPC} is one step shorter than the Supervisor MPC~\eqref{RMPC} in order to maintain the recursive feasibility when switching from the Supervisor MPC to Controller MPC. 
The control input by solving the optimization problem~\eqref{CMPC} is $u^{\text{MPC}}_k = \Tilde{u}^*_{0|k} + K (x_{k} - \Tilde{x}^*_{0|k} )$, where, $K$ is from~\eqref{A_k}.


\section{Recursive feasibility}

\subsection{Terminal set computation}




In order to ensure that the car can always safely overtake obstacles on the road, using the proposed RSCA, we should guarantee that the car can always go as close as possible to one of the road boundaries. 

\vspace*{-3mm}
\begin{definition}
		\normalfont{A \textbf{\textit{safe reference}} is a reference trajectory parallel to one of the road boundaries that the car should be able to reach at the end of every MPC horizon.}
	\end{definition}
\vspace*{-2mm}

\begin{figure}[!]
    \centering
    \includegraphics[scale = 0.5]{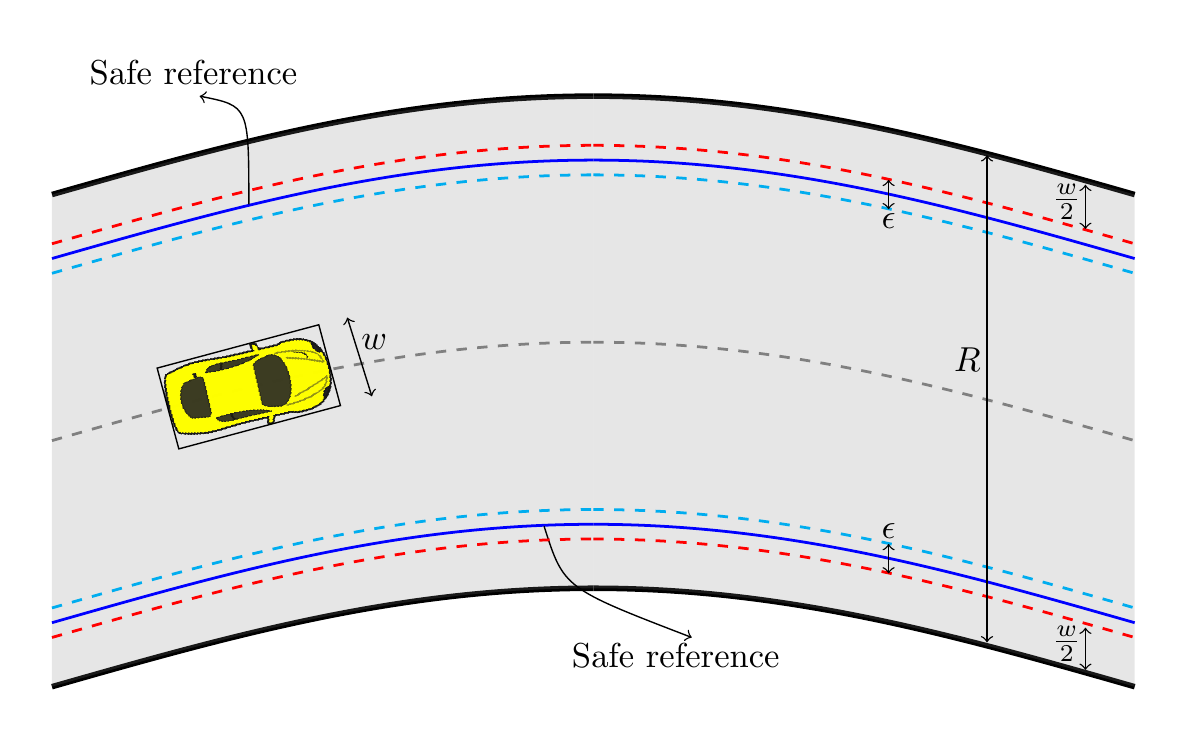}
    \caption{Safe references }
    \label{fig_safe_ref}
\end{figure}

 In \cref{fig_safe_ref}, the red dashed lines located at $\frac{w}{2}$ meters, where $w$ is given in \cref{para-table}, away from the road boundaries represent the safe references which are the last point where the car can go without violating the road boundaries. Enforcing the system to exactly reach the safe reference at the end of every MPC horizon can result in a small terminal set, if one can be found. 
 Therefore, similar to~\cite{Nguyen2021}, we suggest to relax the state constraint by introducing a constant term $\epsilon$, as illustrated in \cref{fig_safe_ref}. Then, the safe references are placed at $\frac{R}{2} - \frac{w}{2} - \frac{\epsilon}{2}$. 

Therefore, two terminal sets can be calculated. One enforces the system to reach the area between the red and cyan dashed lines on the top road boundary. The other one enforces the car to get to the area between the red and cyan dashed lines in the bottom road boundary. Based on of size and the placement of the obstacle, it is decided from which side to overtake. Respectively, one of the terminal sets is used in the optimization problems~\eqref{RMPC} and~\eqref{CMPC}. In the calculation of a terminal set at the upper road boundary, the state constraint on $e^{\rm y}_k$ is $\frac{R}{2} - \frac{w}{2} - \epsilon \leq  e^{\rm y}_k \leq \frac{R}{2} - \frac{w}{2}$. Similarly, for the lower road boundary, it is $ -  \frac{R}{2} + \frac{w}{2}  \leq  e^{\rm y}_k \leq -\frac{R}{2} + \frac{w}{2} + \epsilon $. Then, the state constraints, $\mathcal{X}$, for computing the terminal set can be formulated in polytopic form. 


Before computing the terminal sets, the system coordinates should be transformed from the center-line of the road to the safe reference. If the system coordinates are going to be transformed to the top safe reference, a constant vector $x^{\text{sr}}$ is defined as $x^{\text{sr}}  = \begin{bmatrix} \frac{R}{2} - \frac{w}{2} - \frac{\epsilon}{2} & 0 & 0 & 0 \end{bmatrix}^\top$. If the system coordinates are going to be transformed to the bottom safe reference, $x^{\text{sr}}$ is defined as $x^{\text{sr}} = \begin{bmatrix} -\frac{R}{2} + \frac{w}{2} + \frac{\epsilon}{2} & 0 & 0 & 0 \end{bmatrix}^\top$. Considering $q_{k} = x_{k} - x^{\text{sr}}$, the state space model~\eqref{dis_model_with_disturbance} is rewritten as follows:
\begin{multline}\label{updating_the_model}
           q_{k+1} + x^{\text{sr}} = A ( q_{k} + x^{\text{sr}}) + B u_{k}+ E \dot{\psi}^{\rm{des}}_{k} + d_k, \\
    q_{k+1} = A q_{k} + B u_{k} + E \dot{\psi}^{\rm{des}}_{k} + d_k + (A - I )x^{\text{sr}},
\end{multline}
where, $q_{k} \in \mathbb{R}^4$ is the shifted system vector. Although $\dot{\psi}^{\rm{des}}_k$ changes over the road, we want $\mathcal{X}_N$ to be used for all road curvatures $\dot{\psi}^{\rm{des}}_k \in \Psi$. Therefore, in the terminal sets computations, the term $E \dot{\psi}^{\rm{des}}_{k}$ is also considered as an unknown part, of which only the upper and lower bounds are available. Then, by defining a new disturbance term as $\Tilde{d}_k = E \dot{\psi}^{\rm{des}}_{k} + d_k + (A - I )x^{\text{sr}}$, the state space model is reformulated as:
\begin{equation}
    q_{k+1} = A q_{k} + B u_{k} + \Tilde{d}_k.
        \label{linear_model_after_transformation}
\end{equation}
The new disturbance set is $\tilde{\mathcal{D}} = \mathcal{D} \oplus E \Psi \oplus (A - I )x^{\text{sr}}$ and $\Tilde{d}_k \in \tilde{\mathcal{D}}$. The state constraint for system~\eqref{linear_model_after_transformation} is $q_k \in \mathcal{Q}$, where $\mathcal{Q} = \mathcal{X}  \ominus x^{\text{sr}}$. By using the linear controller $u_k = K q_k$ with $K$ from \eqref{A_k}, the system \eqref{linear_model_after_transformation} is transformed into:
\begin{equation}\label{closed_loop_sys}
    q_{k+1} = A_{\text{K}} q_{k} + \Tilde{d}_k.
\end{equation}

\vspace*{-4mm}
\begin{definition}[\cite{Blanchini1999}]
		\normalfont{The set $\Omega \in \mathbb{R}^n$ is said to be a \textbf{\textit{robust positively invariant set}} of the system~\eqref{closed_loop_sys} if $q_{k+1} \in \Omega$ for all $ q_{k} \in \Omega$ and all $ \Tilde{d}_k \in \tilde{D}$.}
\end{definition}
\vspace*{-3mm}
There are many papers on the computation of $\Omega$ which can be used, e.g.,~\cite{Rakovic2005,Nguyen2014}. However, $\Omega$  cannot directly be used in the Supervisor and Controller MPC. Because it is designed for the transformed system. Therefore, it should be shifted back to the original coordinate system, i.e., $\mathcal{X}_N = \Omega \oplus x^{\text{sr}}$.



\subsection{Proof of recursive feasibility}


At first we prove that at step $k$ when a detection event happens, $u^{\rm{backup}}_k$ satisfies the input constraint and that by applying $u^{\rm{backup}}_k$ to the system~\eqref{dis_model_with_disturbance}, the system state in the next step will satisfy the state constraint. The proofs depend on the calculation of $\mathcal{Z}$. To reduce the conservativeness of the TMPCs, it is desirable to have a smaller $\mathcal{Z}$. Thus, in \cref{neglecting_additive_term}, we show that since $\dot{\psi}^{\rm{des}}_{k}$ is known in every step, it can be neglected in the computation of $\mathcal{Z}$.
\vspace*{-2.5mm}
\begin{proposition}\label{neglecting_additive_term}
\normalfont{
Consider system~\eqref{dis_model_with_disturbance} and the nominal system~\eqref{dis_model}.
Next, the control input $u_k$ of the system~\eqref{dis_model_with_disturbance} is taken as follows:
\begin{equation}
\label{eq:input_to_prop_system}
    u_k= \bar{u}_k + K(x_k - \bar{x}_k),
\end{equation}
with $K$ the control gain from~\eqref{A_k}. Define $\mathcal{Z}$ as the disturbance invariant set for the system $x_{k+1}=A_{\text{K}} x_k + d_k$ which satisfies:
\begin{equation}
\label{eq:disturbance_set_property}
    A_{\text{K}} \mathcal{Z} \oplus \mathcal{D} \subseteq \mathcal{Z}.
\end{equation}
If $x_k \in \bar{x}_k \oplus\mathcal{Z}$, then $x_{k+1} \in \bar{x}_{k+1} \oplus \mathcal{Z}$.}
\end{proposition}
\vspace*{-2mm}
\begin{proof}
By introducing $e_k \triangleq x_k - \bar{x}_k$, we have that $e_{k+1}= A_{\text{K}} e_k + d_k $. Then, since $e_k \in \mathcal{Z}$ (i.e., $x_k \in \bar{x}_k \oplus \mathcal{Z}$) and $ A_{\text{K}} \mathcal{Z} \oplus d_k \subseteq \mathcal{Z}$ as required in~\eqref{eq:disturbance_set_property}, we have that $e_{k+1}\in \mathcal{Z}$ which results in $x_{k+1} \in \bar{x}_{k+1} \oplus \mathcal{Z}$. 
\end{proof}
\smallskip
In the Supervisor MPC problem \eqref{RMPC}, the initial condition is the predicted nominal state at the next step \eqref{estimated_state}, not the actual system state \eqref{real_state}. Therefore, to ensure that the backup control input \eqref{take_over_control_input} and its application to the system \eqref{dis_model_with_disturbance} are safe, just considering the disturbance of one step in the computation of $\mathcal{Z}$ is not enough, but the propagation of the disturbance in two steps should be considered. This brings us to the following corollary.


\vspace*{-2.5mm}
\begin{corollary}\textbf{(Safety of the vehicle at the moment of detection event)}
\label{corollary}
\normalfont{ Define a set $\mathcal{Z}$ as used in~\eqref{RMPC} and~\eqref{CMPC} to satisfy the following condition:
\begin{equation}
\label{eq:Z_property}
    A_{\text{K}} \mathcal{Z} \oplus \mathcal{D} \oplus A_{\text{K}} \mathcal{D} \subseteq \mathcal{Z},
\end{equation}
in which, $A_{\text{K}}$ is from~\eqref{A_k}. Then, the control input $u^{\rm{backup}}_k$ in~\eqref{take_over_control_input} satisfies the input constraint. Furthermore, if $u^{\rm{backup}}_k$ is applied to the real system~\eqref{dis_model_with_disturbance} at step $k$, then, the next state $x_{k+1}$ is still within the state constraint set $\mathcal{X}_{k+1}$ regardless the value of the disturbance, $d_k$.}
\end{corollary}
\vspace*{-2mm}
\begin{proof} This proof consists of two parts. The first part is on proving that $u^{\rm{backup}}_k \in \mathcal{U}$. From~\eqref{real_state}, having that $x_k = \hat{x}_k + d_{k-1}$ leads to:
\begin{align*}
    u^{\rm{backup}}_k &= \bar{u}^*_{0|k-1} + K(x_k - \bar{x}^*_{0|k-1}) \\
    &= \bar{u}^*_{0|k-1} + K (\hat{x}_k - \bar{x}^*_{0|k-1}) + K d_{k-1}. 
\end{align*}
With~\eqref{intial_condition_in_MPC}, it follows that $\hat{x}_k \in \bar{x}^*_{0|k-1} \oplus \mathcal{Z}$. Thus, $\hat{x}_k - \bar{x}^*_{0|k-1} \in \mathcal{Z}$ holds. With~\eqref{input_contsraint_in_MPC}, it also holds that $\bar{u}^*_{0|k-1} \in ( \mathcal{U} \ominus K \mathcal{Z}) \ominus K \mathcal{D}$. 
Therefore:
\begin{equation*}
    u^{\rm{backup}}_k \in (\mathcal{U} \ominus K \mathcal{Z} ) \ominus K \mathcal{D}   \oplus K \mathcal{Z} \oplus K \mathcal{D} \subseteq \mathcal{U}.
\end{equation*}

Next, it is shown that by applying $u^{\rm{backup}}_k$ to the system~\eqref{dis_model_with_disturbance} the state constraint is satisfied in the next step $k+1$:
\begin{equation}\label{proof_of_X_safety}
\begin{aligned}
x_{k+1} ={} & Ax_k + B (\bar{u}^*_{0|k-1} + K(x_k - \bar{x}^*_{0|k-1})) \\
      & + E \dot{\psi}^{\rm{des}}_{k} + d_k. \\
\end{aligned}
\end{equation}

Because $x_k = \hat{x}_k + d_{k-1}$ from~\eqref{real_state}, equation~\eqref{proof_of_X_safety} is equivalent to $x_{k+1} = A (\hat{x}_k + d_{k-1}) + B (\bar{u}^*_{0|k-1} + K(\hat{x}_k + d_{k-1} - \bar{x}^*_{0|k-1})) + E \dot{\psi}^{\rm{des}}_{k} + d_k,$ which leads to:
\begin{equation}\label{dis_system_col}
\begin{aligned}
x_{k+1} ={} & A \hat{x}_k + B\bar{u}^*_{0|k-1} + BK(\hat{x}_k - \bar{x}^*_{0|k-1}) + \\
  &E \dot{\psi}^{\rm{des}}_{k} + d_k + A_{\text{K}} d_{k-1}.
\end{aligned}
\end{equation}
The nominal system~\eqref{dis_model}, after applying $\bar{u}^*_{0|k-1}$ is as follows:
\begin{equation}\label{nominl_system_col}
    \bar{x}_{1|k-1}^* = A \bar{x}_{0|k-1}^* + B\bar{u}^*_{0|k-1} + E \dot{\psi}^{\rm{des}}_{k} \in \bar{\mathcal{X}}_{1|k-1} \ominus \mathcal{Z}. 
\end{equation}
 
By applying \cref{neglecting_additive_term} to the two systems~\eqref{dis_system_col} and~\eqref{nominl_system_col}, i.e. considering $w \triangleq d_k + A_{\text{K}} d_{k-1}$, and using the property of $\mathcal{Z}$ as in~\eqref{eq:Z_property}, we have:
\begin{equation}
\label{eq:result_of_corollary}
    x_{k+1} \in \bar{x}_{1|k-1}^* \oplus \mathcal{Z},
\end{equation}
since $\hat{x}_k \in \bar{x}_{0|k-1}^* \oplus \mathcal{Z}$ as constrained in ~\eqref{intial_condition_in_MPC}. Note that, we have $\bar{x}_{1|k-1}^*\in \mathcal{X}_{k+1} \ominus \mathcal{Z}$ as required from~\eqref{state_constraint_in_MPC} which leads to $x_{k+1} \in \mathcal{X}_{k+1} \ominus \mathcal{Z} \oplus \mathcal{Z} \subseteq \mathcal{X}_{k+1}$.
\end{proof}

\vspace*{-4mm}
\begin{theorem} \textbf{(Recursive feasibility)}
\normalfont{
The feasibility of the Supervisor MPC at step $k-1$ guarantees the feasibility of the Controller MPC at step $k+1$.}
\end{theorem}
\vspace*{-2mm}
\begin{proof}
Feasibility of the Supervisor MPC at step $k-1$ provides the following optimal state and input sequences:
\begin{align*}
    \bar{x}^*_{0|k-1}, \bar{x}^*_{1|k-1}, \cdots, \bar{x}^*_{N|k-1}, \\
    \bar{u}^*_{0|k-1},  \bar{u}^*_{1|k-1}, \cdots,  \bar{u}^*_{N-1|k-1}. 
\end{align*}

From \cref{corollary}, we have $x_{k+1} \in \bar{x}_{1|k-1}^* \oplus \mathcal{Z}$ as in~\eqref{eq:result_of_corollary} and hence, $\bar{x}_{1|k-1}^*$ is a feasible candidate for $\Tilde{x}_{0|k+1}$. Next, since $\bar{u}^*_{1|k-1} \in (\mathcal{U} \ominus K\mathcal{Z}) \ominus K \mathcal{D} \subseteq \mathcal{U} \ominus KZ$, it can be used as a feasible candidate for $\Tilde{u}_{0|k+1}$. This leads to $\bar{x}^*_{2|k-1} = A \bar{x}^*_{1|k-1} + B \bar{u}^*_{1|k-1} + E \dot{\bar{\psi}}_{1|k-1} \in \bar{\mathcal{X}}_{2|k-1} \ominus \mathcal{Z} \equiv \mathcal{X}_{k+2} \ominus \mathcal{Z}$, which is a valid candidate for $\Tilde{x}^*_{1|k+1}$.


By doing the same process, it will be clear that the sequence of $\bar{x}^*_{1|k-1}, \cdots, \bar{x}^*_{N|k-1}$ and $\bar{u}^*_{1|k-1}, \cdots,  \bar{u}^*_{N-1|k-1}$, are valid candidates for the Controller MPC at step $k+1$. The recursive feasibility before or after a detection event follows the standard proof in the literature, i.e.,~\cite{Mayne2005}.
\end{proof}

\section{Simulation Results} 

To showcase the applicability of the proposed approach, 120 different obstacle avoidance scenarios are simulated using MATLAB~\cite{MATLAB:2019} and the results compared with the SCA in~\cite{Nezami2021}. 
In the MPC setup, the sampling time $t_{\rm s}$ is $0.1$ seconds. The tuning parameters are $Q = \text{diag}(\begin{bmatrix} 1 & 1 & 1 & 1  \end{bmatrix})$, $R = 0.1$, and $N = 30$. 

In the simulations, the operating controller is a Pure Pursuit (PP) controller which is a path-tracking algorithm~\cite{Woods2013}. The PP controller tries to keep the car on the reference trajectory by minimizing the car's deviation from the reference trajectory. The look-ahead time is chosen as $0.5$ seconds.

Model~\eqref{dis_model_with_disturbance} represents the actual lateral dynamics of the car in the simulations, while the nominal model~\eqref{dis_model} is used in the Supervisor and Controller MPCs. The vehicle's parameters used in the simulation are shown in \cref{para-table}. For the calculation of $\mathcal{Z}$, the approach in~\cite{Rakovic2007} is used. The calculation of $\mathcal{X}_N$ is carried out in accordance with~\cite[p.~25]{Nguyen2014}, with $\epsilon = 0.5 \rm m$ for initializing $x^{\text{sr}}$ in~\eqref{updating_the_model}.

The constraint on $|e^{\rm y}_k|$, for the steps over the prediction horizon where there is no obstacle is enforced in the optimization problems as $|e^{\rm y}_k|\leq \frac{R}{2} - \frac{w}{2}$, where $\frac{ R}{2} = 8 \rm m$ and $ w$ is given in \cref{para-table}. Otherwise, the constraint is updated based on the size of the obstacle, e.g., for a scenario, where an obstacle with a width of $\text{obs}^{\rm width}$ is located at the center of the road, the constraint is $\frac{\text{obs}^{\rm width}}{2} + \frac{w}{2} \leq e^{\rm y}_k\leq \frac{R}{2} - \frac{w}{2}$. The constraints on the rest of the states are $|\Delta e^{\rm y}_k|\leq 10  \rm m/\rm s$, $|e^{\rm \psi}_k|\leq \frac{\pi}{2} \rm rad$ and $|\Delta e^{\rm \psi}_k|\leq \frac{\pi}{3 t_{\rm s}} \rm rad/ \rm s$. The input is also bounded by $|u_k| \leq \frac{34 \pi}{180} \rm rad$. 

\begin{figure}[!]
\input{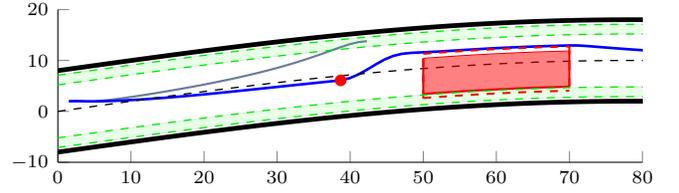}
\caption{A safe obstacle avoidance maneuver by RSCA}
\label{fig_safe_manuver}
\end{figure}

In \cref{fig_safe_manuver}, a safe obstacle avoidance scenario is illustrated. In this scenario, the centerline of the road is the reference trajectory, and the blue line indicates the car's movement. The obstacle is $50 \rm m$ away from where the car starts. The car moves at a speed of $12 \rm m/ \rm s$. In this scenario, the PP controller drives the car right to the obstacle, and the Supervisor MPC should intervene when necessary. A detection event takes place in the Supervisor MPC at $11.3 \rm m$ before the obstacle. This point is shown with a red dotted line. Thereafter, the Controller MPC takes over the car, and the car safely overtakes the obstacle. The gray line in the \cref{fig_safe_manuver} shows the optimal state sequence, generated by the Supervisor MPC at step $k$, over the entire prediction horizon. As evident in~\cref{fig_safe_manuver}, the car ends up in a safe reference at the end of the horizon.

To evaluate the RSCA and compare it with the SCA, 120 obstacle avoidance scenario simulations are carried out. 
The obstacles widths are chosen randomly based on uniform distribution within the range $0.1$ and $2.5\rm m$ and for the length between $1$ to $10 \rm m$ as well as the vehicle speed between $5$ to $20 \rm m/ \rm s$. Between these $120$ scenarios, one third of them is carried out with a disturbance set of $|d^1_k| \leq \boldsymbol{1} \cdot 10^{-2}$, one third with $|d^2_k| \leq \boldsymbol{1} \cdot 10^{-3}$, and 40 scenarios with $|d^3_k| \leq \boldsymbol{1} \cdot 10^{-4}$. In the simulations, the disturbances are randomly generated based on the uniform distribution according to the assigned ranges. The initial conditions are selected such that the MPC at $k=0$ is feasible.  
 
The measure of success or failure in each scenario is the feasibility or infeasibility status of the Controller MPC after a \emph{detection event}. In $31\%$ of all scenarios, the SCA failed in performing a safe obstacle avoidance scenario, while the RSCA succeeded in all of them. The chance of failure of the SCA was higher in more complex scenarios, e.g., if the obstacle's dimension was larger. When comparing the intervention time of both control approaches, it is noticeable that in some scenarios the RSCA reacts before the SCA. In total in $13\%$ of the scenarios, the RSCA intervened the operating controller one or two sampling times earlier than the SCA. Most of the scenarios with the earlier intervention by RSCA belonged to the scenarios with larger disturbance $d^1_k$. In total, $75\%$ of the earlier interventions by the RSCA are related to the set $d^1_k$ of larger disturbances, while $19\%$ and only $6\%$ is the share of $d^2_k$ and $d^3_k$, respectively. The earlier intervention of the RSCA compared to the SCA implies that the robust Supervisor is more conservative than the standard  MPC in the SCA. However, the relation of the size of the expected disturbance and the conservativeness of the RSCA could be exploited to adapt to the situation, e.g. by considering the current uncertainty in the system states. Furthermore, the more conservative controller is acceptable, since the RSCA was capable to perform a safe obstacle avoidance maneuver in all scenarios.

\section{CONCLUSIONS}
The implementation of autonomous driving in real world applications requires rigorous satisfaction of safety-critical constraints in an uncertain environment. Thus, an RSCA is proposed to include disturbances acting on the system while guaranteeing constraint satisfaction during obstacle avoidance maneuvers independent of the roads’ curvature.
The proposed RSCA utilized two TMPCs, one for predicting a potential safety event, and one for taking over from the unsafe operating controller after a detection event. The general benefits of RSCA compared with~\cite{Nezami2021} were displayed in a case study of $120$ different scenario simulations.


\printbibliography

\addtolength{\textheight}{-12cm}   

\end{document}